# Simple Wideband RCS Reduction by Phase Gradient Modulated Surface

Yousef Azizi, Mohammad Soleimani, Seyed Hasan Sedighi, and Ladislau Matekovits, *Senior Member, IEEE*

*Abstract*—This paper presents the design and implementation of a simple, single-layer, broadband (97%, 11.3-32.3 GHz) Radar Cross Section Reduction (RCSR) Modulated Surface (MS). It uses modulation of the edge-length of the square patch (SP) radiators within adjacent unit cells. By using Sinusoidal Modulation (SM) of the edge length of the unit cells, the unit cells sequences with phase gradient, that plays an effective role in improving the RCSR, can be used for wideband RCSR achievement. The proposed structure with the dimension of 250×250mm$^2$ that consists of 40 × 40 unit cells with period of 6mm printed on a RO4003 substrate of 1.6mm thickness and has been considered. Measurements on a prototype were conducted considering both mono- and bi-static arrangements for oblique incidences for both TM and TE polarization tests. A good agreement between simulation and measurement results proves the validity of the design criteria.

## I. INTRODUCTION

Metasurfaces are among the important structures in RCSR that are widely used today. Amplitude and phase of the reflection are the main design factors of metasurfaces. Using artificial magnetic conductor (AMC) unit cells consisting of N×N arrays (called tile), gives rise to formation of different reflection phases for them, hence opening the way to phase cancellation and consequently to RCSR. By using two tiles with 180°±37° reflection phase difference, 61% RCSR bandwidths has been reported in [1]. By implementation of dual and triple band AMCs with different resonances, 85% and 91.5% RCSR bandwidth was reported in [2] and [3], respectively. In [4] and [5], by using multilayer structures with higher complexity RCSR bandwidth further increased up to 109% and 109.4%, respectively. Pixelated checkerboard structure with 95% monostatic RCSR bandwidth was reported in [6] that suffer of thick substrate (6mm). By varying the surface substrate thickness, which is however difficult from practical point of view and costly, more suitable phase gradient can be created between adjacent tiles, and structures with 113%, 148%, and 122.3% RCSR bandwidths were reported [7-9]. Use of metasurface in random configuration also resulted in 73% and 77% RCSR bandwidth in [10] and [11], respectively. Using holographic surfaces to convert the impinging wave into surface one, as well as anomalous or diffuse reflection of the incidence, RCSR can be achieved as reported for example in [12-14]. Due to the narrow bandwidth of the holographic surface, PGM structures with anomalous reflection in [15] and [16] reports 61.2% and 82.4% RCSR bandwidth, respectively. In these structures, more than two unit cells with a phase difference of less than 180°±37° were used to improve RCSR characteristics. Using a large number of unit cells with phase gradient and low reflection amplitude arranged in a SM configuration allowed generation of a surface with more suitable phase gradient and less amplitude reflection to be intelligently placed with more repetitions in the structure and RCSR bandwidth was effectively increase to 128% [17].

In this paper, a SP unit cell with non-uniform gradient reflection phase by changing the gap size is perceptively placed in the structure giving rise to sinusoidal modulation so that wideband RCSR bandwidth achieved. It is shown that the using of non-identical number of unit cells, which is easily possible by sinusoidal modulation, leads to RCSR bandwidth enhancement. Finally, the measurement results of a fabricated prototype in mono-static, bi-static and for oblique incidence (15°-40°) and for TM/TE polarizations are presented. Measurement results indicate 97% of the 10-dB RCSR bandwidth for normal incidence and 54% and 45% for TM and TE incidences for the maximum considered angle, respectively. Low profile (1.6mm substrate thickness), simple structure and wideband RCSR performance are the main advantage of proposed structure compare with state of arts refrences.

The paper is structured as follows: design concept and parametric analysis are presented in Section II. Identification of best configuration has been prototyped and measurements results compared with the numerical data are presented in the Section III. The final section is devoted to the conclusions.

Y. Azizi and M. Soleimani are with the Department of Electrical Engineering, Iran University of Science and Technology (IUST), Tehran, Iran (e-mail: yousefazizi@elec.iust.ac.ir; soleimani@iust.ac.ir).

S. H. Sedighy is with the School of New Technologies, Iran University of Science and Technology, Tehran, Iran (e-mail: sedighy@iust.ac.ir)

L. Matekovits is with the Department of Electronics and Telecommunications, Politecnico di Torino, Corso Duca degli Abruzzi 24, I-10129 Turin, Italy (e-mail: ladislau.matekovits@polito.it). He is also with Istituto di Elettronica e di Ingegneria dell'Informazione e delle Telecomunicazioni, National Research Council of Italy, 10129 Turin, Italy and with Department of Measurements and Optical Electronics, University Politehnica Timisoara, 300006 Timisoara, Romania.

## II. DESIGN CONCEPT

The RCSR value in AMC structures that have two single cells is calculated by the following equation [1]

$$RCSR(dB) = 10\log\left|\frac{|\Gamma_1|e^{j\angle\Gamma_1}+|\Gamma_2|e^{j\angle\Gamma_2}}{2}\right|^2 \quad (1)$$

where $|\Gamma_1|$, $|\Gamma_2|$, $\angle\Gamma_1$ and $\angle\Gamma_2$ are the reflection amplitude and phase from the unit cells of type 1 and 2, respectively.

Usually, the reflection phase is considered as the main design factor, but if both variables (amplitude and phase) are used in RCSR calculation/definition, RCSR bandwidth could be effectively increased. In [17], an ultra-wideband RCSR has been reported by using amplitude and phase gradient MS that has a substrate thickness of 3mm. The high thickness of the substrate as well as the use of stacked SPs have reduced the reflection of the amplitude of the unit cells and efficiently increased the bandwidth of the RCSR. According to Eq. (1), if the reflection amplitude of both unit cells is equal to 1, the phase difference of 180°±37° between the two unit cells causes an RCSR of at least 10 dB. It is relatively challenging to design two single cells that have the desired phase difference in wideband, and therefore, according to [15-18] more unit cells can be used that have a smaller phase difference one with respect to each other. If *n* types of cells are used, where each of them has its own number of repetition ($m_i$) and they totally reflect the incident waves, i.e. $|\Gamma_i| = 1|$, Eq. (1) is expressed as [17]

$$RCSR(dB) = 10\log\left|\frac{m_1 e^{j\angle\Gamma_1} + m_2 e^{j\angle\Gamma_2} + \ldots + m_n e^{j\angle\Gamma_n}}{\sum_1^n m_i}\right|^2 \quad (2)$$

In this equation, the reflection phase and the number of unit cells from each type (*m*i) are the effective variables in RCSR. In fact, the number of each cell as a weighting factor can help increasing the RCSR bandwidth. The reflection phase of the SP unit cell is plotted in Fig.1 for seven different gap values (g) from 0.1mm to 2.9mm.

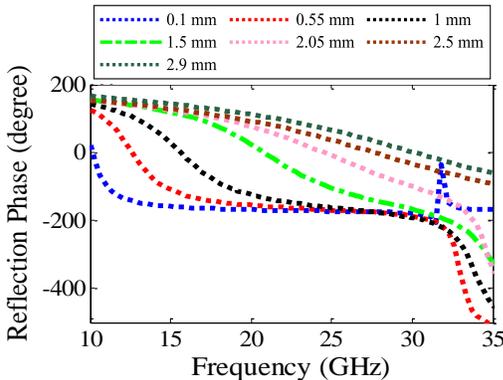

Fig. 1. Reflection phase of the SP unit cells with period equal to 6mm for different gap width *g*.

The SP unit cells was designed on a grounded RO4003 substrate ($\varepsilon_r = 3.55$) with 1.6mm thickness and 6mm period. The used substrate has relatively low losses and its dispersion effects can be ignored, so the reflection amplitude of the SP cells in the frequency range of 10-35 GHz is approximately equal to 1 ($|\Gamma_i| = 1$).

It is observed that the reflection phase variations are not uniform in the whole 10-35 GHz band. For small amounts of gap (0.1-1mm), relatively suitable phase difference between the SP cells is formed at low frequencies (10-20 GHz), where the reflection phase of these cells is same at high frequencies (reflection phase difference very small and could not help for RCSR). On the other hand, with increasing gap (1.5-2.9mm), the desired phase difference between SP cells is formed at higher frequencies (35-35 GHz) and the phase difference of lower frequencies are not satisfying.

However, in some references e.g., [8], [15], and [19], by using coding metasurface, substrate thickness change as well as unit cell rotation, relatively uniform phase differences between unit cells have been reported. Although these structures have more RCSR bandwidth, they are costly and complex. In fact, the existence of the equal phase difference between the unit cells has a more effective role in designing the wideband RCSR structures, and if the phase differences between all unit cells are nearly equal to each other, the maximum RCSR bandwidth can be achieved. However, when the phase difference between the unit cells is not same as plotted in Fig.1, selecting the none equal weight coefficients ($m_i$ in (2)) for all the SP unit cells can be helpful in wideband RCSR achievement.

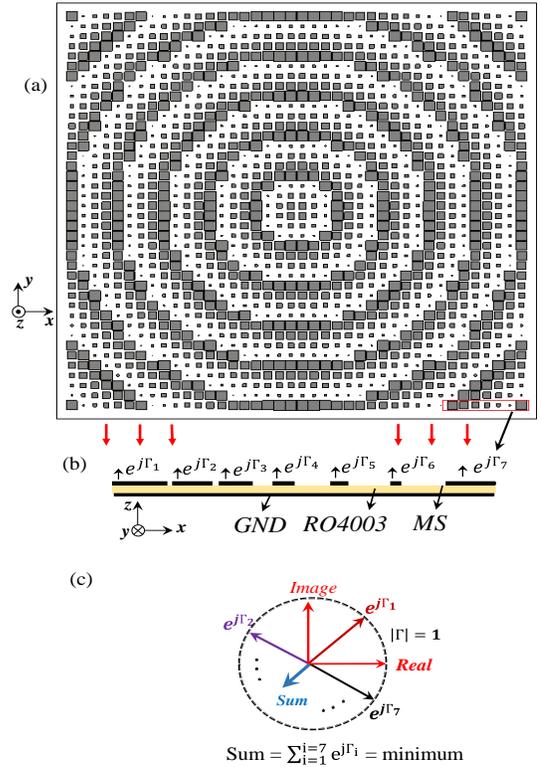

Fig. 2. Schematic of PGM structure and RCSR mechanism: (a) PGM structure, (b) PGM surface performance when illuminated by incidence plan wave, (c) RCSR mechanism of PGM by phase cancellation.

Figure 2(a) shows a schematic of the proposed structure that consists of SP unit cells (with dimensions of Fig. 1) arranged in sinusoidal modulation to obtain wideband RCSR structure

Figure 2(b) shows the reflection performance of a part of the structure when illuminated by an incident plan wave. By using a 1.6mm thick RO4003 substrate characterized by low loss causes the complete reflection of the entire incidence wave. Therefore, the reflection phase as the main variable for RCSR in the form phase cancellation (Fig. 2(c)) increases the RCSR bandwidth. SP unit cells with gap size (*g*) of 0.1, 0.55, 1, 1.55, 2.05, 2.5 and 2.9mm was repeated with the *m*i (i = 1, …, 7) *coefficient of 104, 112, 164,196, 412, 336 and 276, respectively*. These coefficients are the optimal values that obtained by Genetic Algorithm (GA) that causes (2) to be minimized. In GA procedure, the reflection phase of SP unit cells (Fig.1) considered as an inputs and *m*i (coefficient of each unit cell) considered as an unknown parameter that potentially lead the Err function to be minimized. After implement of GA the Err function ((2)) with the above mentioned coefficients (*m*i) was minimized. In order to implement the structure, the CST software is linked with the MATLAB such that the SP unit cells arranged in MS form lead to maximum RCSR achievement Then the introduced structure with dimensions of 250 × 250mm$^2$ composed of 1600 unit cells was achieved that has a RCSR bandwidth of 97% (11.3-32.3 GHz).

## III. MEASUREMENT AND RESULTS

Figure 3 shows the prototype of the proposed broadband RCSR structure that is printed on a RO4003 substrate with dimension of 250×250mm$^2$.

RCS measurements have been carried out in an anechoic chamber using N5227A PNA network analyzer as a source and receiver. The 30 GHz test bandwidth (10-40 GHz) is covered by three sets of Tx/Rx antenna. The test method is based on time gating and it needs post processing of the initial results to extract the final values. Two antennas are used to transmit and receive the chirp signals of the PNA as reported in Fig. 3.

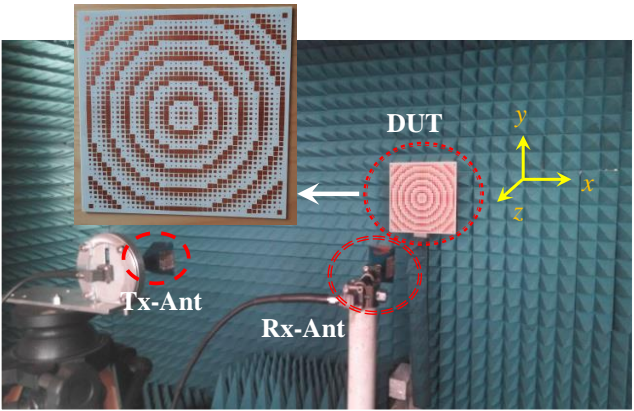

Fig. 3. Fabricated structure and RCS test set-up

The measured result of the monostatic RCSR structure shown in Fig. 4 is in good agreement with the simulation one, which shows the successful performance of the design method as well as the post-processing after time-gating method. Based on the measured results in the frequency range of 11.3-32.3 GHz (97%), RCSR more than 10-dB is achieved. Comparison between simulation and measurement results of Fig. 5 shows that by using SM and applying intelligent weighting coefficients in a SP unit cell, wideband RCSR performance can be achieved.

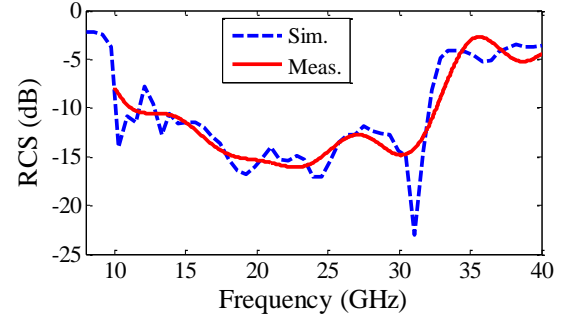

Fig. 4. Monostatic RCS results of the PGM structure

The use of SP unit cells with different gap dimensions and their placement in a MS configuration in association with applied optimal weight coefficients to each cell has induced diffuse reflection from the surface of the structure. The RCS pattern of the structure at 18 GHz and 31.1 GHz, shown in Fig. 5 shows that a large part of the incident plane wave is redirected in different directions by scattering the input wave.

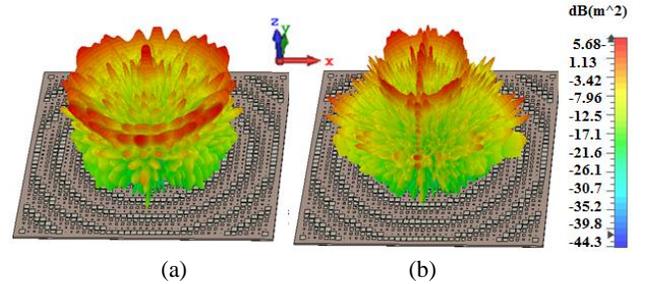

(a) (b)
Fig. 5. RCS pattern of the PGM structure: (a) 18 GHz, (b) 31.1 GHz.

As shown in Fig. 5, at 18 GHz and 31.1 GHz the structures redirects the reflection: most of the redirection main lobes occurs at 45° and 24° angles, respectively, which can be clearly seen in the Cartesian plots in Fig.6. Figure 6 (a, b) show a 2D RCS measurement and simulation patterns of the structure at 18 and 31.1 GHz, respectively. It should be noted that due to equipment limitations, the measurement of the bi-static RCS pattern are done with an angular step of 1° and therefore there are some rapid changes in the measurement results due to this issue while the presented simulation results have an accuracy of 0.25 degrees. As shown in Fig. 6(a), the 2D RCS pattern is perfectly symmetrical, and at 18 GHz at angles of 1°-60°, there is good agreement between the simulation and measurement results. But at 70° to 90° angles there are some inconsistencies that can be caused by manufacturing errors. It is noteworthy that the signal strength at these angles is about 10 dB less than that of the main lobes (0° and 45°), and this radiation is not very effective.

Similarly, as shown in Fig. 6(b), at 31.1 GHz, there is a good agreement between the measurement and simulation results at all angles from 1° to 90°. Although the oscillation rate in the pattern increases with increasing frequency, but the overall measurement results are in good agreement with the simulation.

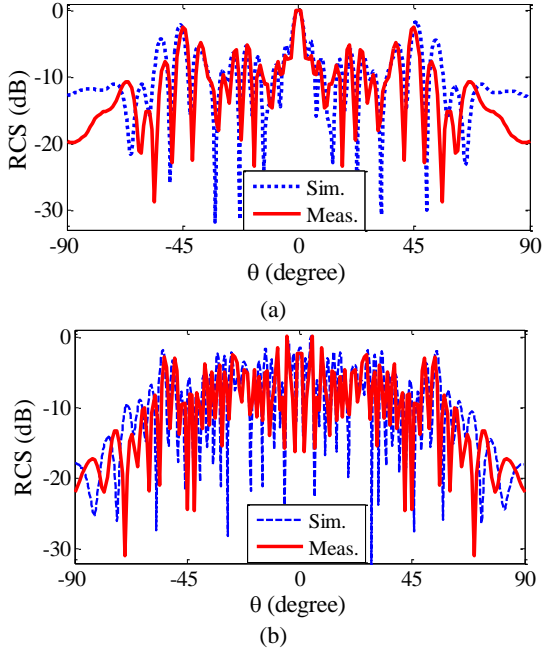

Fig. 6. Bi-static RCS simulation and measurement results: (a) 18 GHz, (b) 31.1 GHz

Another feature of the proposed wideband RCSR MS structure is its relatively low sensitivity to oblique incidence of TM and TE polarizations for oblique incidence up to 40°.

The RCS results for oblique incidence of 15°-40° are plotted in Fig. 7(a-b) for TM and TE polarizations, respectively. Since the reflection phase of the SP unit cell for oblique incidence is more sensitive to TE polarization, it is expected that the proposed structure will be more sensitive to TE polarization and its RCSR bandwidth be limited by increasing the angle of the oblique incidence [21]. By increasing the incidence angle from 15° to 40° in TE polarization, the RCSR bandwidth of the structure limited up to 45% (21.7-34.5 GHz). On the other hand, because the reflection phase of the SP unit cell is less sensitive to the oblique incidence in TM polarization, so by increasing the oblique incidence up to 40°, the RCSR bandwidth is limited up to 54% (17.5-30 GHz).

Comparison between proposed wideband RCSR MS and the state of the art references is presented in Tab. 1. Although [4-5, 7-9, 17, 20], have RCSR bandwidth more than the proposed structure, they are complex and costly to build due to the use of substrates with different thicknesses and also they are multilayer structures. Simplicity of design (single layer), no need for spacers and lump elements, wideband RCSR (97%) and low substrate thickness (1.6mm) are the advantage of the proposed structure.

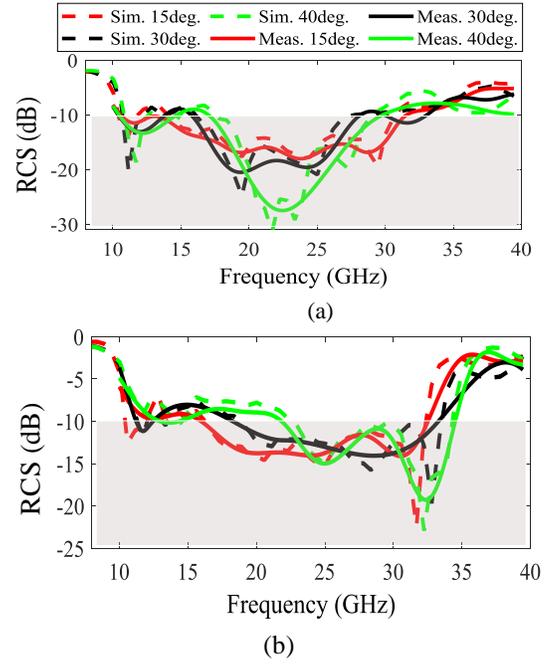

Fig. 7. Sensitivity study of the proposed PGM for oblique incidence: (a) TM polarization, (b) TE polarization

TABLE I
COMPARISON BETWEEN THE PROPOSED PGM AND THE STATE OF THE ARTS

| Structure | Thickness (mm) | BW (%) Freq. (GHz) | No. of layers | Substrates |
|---|---|---|---|---|
| [3] | 11.5 | 91.5/3.77-10.14 | 2 | RO4350B-Air |
| [4] | 2.5 | 109/13.1-44.5 | 3 | FR4/air |
| [5] | 6 | 109/4.8-16.4 | 2 | FR2-Air |
| [7] | 6 | 113/6.4-44.5 | 2 uneven layers | F4B |
| [8] | 6 | 148/6.16-41.3 | 3 uneven layers | F4B |
| [9] | 6 | 122/6.2-25.7 | 3 uneven layers | F4B |
| [16] | 3 | 82.4/7-16.8 | 1 | RO5880 |
| [17] | 3.2 | 128/9-40.7 | 2 | FR4 |
| [19] | 1.5 | 91/15-40 | 1 | F4B |
| [20] | 2.5 | 108/13.6-45.5 | 2 | FR4-FR4 |
| This work | 1.6 | 97/11.3-32.3 | 1 | RO4003 |

IV. CONCLUSION

In this paper, a wideband (97%) RCSR MS is presented, which is of reduced complexity, single-layer and has low sensitivity to inclined illumination for both TM and TE polarizations. In the proposed structure, sinusoidal modulation is used that in addition to creating symmetry in the RCS patterns, the unit cells with suitable phase gradient can be used in greater numbers compare the other cells and the RCSR bandwidth can be increased compared to the case that number of unit cells are equal to each other . Also, with the aim of validating the design, an example of the proposed structure is measured in cases of mono-static, bi-static for oblique incidence for both TM and TE polarizations. The presented experimental results are consistent with the simulation results and confirm the validity of the design method.

## References

[1] W. Chen, C. A. Balanis, and C. R. Birtcher, "Checkerboard EBG surfaces for wideband radar cross section reduction," *IEEE Trans. Antennas Propag.*, 63(6), 2636–2645, 2015.

[2] S. H. Esmaeli and S. H. Sedighy, "Wideband radar cross-section reduction by AMC," *Electronics Letters*, vol. 52, no. 1, pp. 70-71, 2016.

[3] D. Sang, Q. Chen, L. Ding, M. Guo, and Y. Fu, "Design of checkerboard AMC structure for wideband RCS reduction," *IEEE Trans. Antennas Propag.*, vol. 67, no. 4, pp. 2604–2612, Apr. 2019.

[4] E. Ameri, S. H. Esmaeli, and S. H. Sedighy, "Ultra wide band radar cross section reduction using multilayer artificial magnetic conductor metasurface," *J. Phys. D Appl. Phys.* 51(28), 285304 2018.

[5] Y. Zheng, X. Cao, J. Gao, H. Yang, Y. Zhou, and S. Wang, "Shared aperture metasurface with ultra-wideband and wide-angle low-scattering performance," Opt. Mater. Express, vol. 7, no. 8, pp. 2706–2714, 2017.

[6] M. J. Haji-Ahmadi, V. Nayyeri, M. Soleimani, and O. M. Ramahi, "Pixelated checkerboard metasurface for ultra-wideband radar cross section reduction," *Sci. Rep.*, vol. 7, no. 1, pp. 11437, 2017.

[7] J. Su, Y. Cui, Z. Li, Y. Yang, Y. Che, and H. Yin, "Metasurface base on uneven layered fractal elements for ultra-wideband RCS reduction," *AIP Adv*. 8(3), 035027 (2018).

[8] J. Su, H. He, Y. Lu, H. Yin, G. Liu, and Z. Li, "Ultra-wideband radar cross section reduction by a metasurface based on defect lattices and multiwave destructive interference," *Phys. Rev. Appl.* 11(4), 044088, 2019.

[9] J. Su, H. He, Z. Li, Y. Yang, H. Yin, and J. Wang, "Uneven-layered coding metamaterial tile for ultra-wideband RCS reduction and diffuse scattering," *Sci. Rep.* 8(1), 8182, 2018.

[10] Y. Zhuang, G. Wang, J. Liang, T. Cai, X.-L. Tang, T. Guo, and Q. Zhang, ''Random combinatorial gradient metasurface for broadband, wide-angle and polarization-independent diffusion scattering,'' *Sci. Rep.*, vol. 7, Nov. 2017, Art. no. 16560.

[11] P. Su, Y. Zhao, S. Jia, W. Shi, and H. Wang, ''An ultra-wideband and polarization-independent metasurface for RCS reduction,'' *Sci. Rep.*, vol. 6, Feb. 2016, Art. no. 20387.

[12] Y. F. Li et al., "Wideband radar cross section reduction using two-dimensional phase gradient metasurfaces," *Appl. Phys. Lett.* 104(22), 221110, 2014.

[13] Y. Liu, Y. Hao, K. Li, and S. Gong, "Wideband and polarization independent radar cross section reduction using holographic metasurface," *IEEE Antennas Wireless Propag. Lett.*, vol. 15, pp. 1028–1031, Mar. 2016.

[14] Y. Azizi, M. Soleimani, and S. H. Sedighy, "Low cost, simple and broad band radar cross section reduction by modulated and holography metasurfaces," *J. Phys. D Appl. Phys.* 52(43), 435003, 2019.

[15] Q. Zheng et al. "Wideband, wide-angle coding phase gradient metasurfaces based on Pancharatnam-Berry phase," *Sci. Rep.* 7, 43543, 2017.

[16] W. Zhang, Y. Liu, S. Gong, J. Wang, and Y. Jiang, "Wideband RCS reduction of a slot array antenna using phase gradient metasurface," *IEEE Antennas Wireless Propagation Lett.* 17(12), 2193–2197, 2018.

[17] A, Yousef, M. Soleimani, and S. H. Sedighy, "Ultra-wideband radar cross section reduction using amplitude and phase gradient modulated surface," *Journal of Applied Physics* 128, no. 20, pp, 205301, 2020.

[18] T. Song, L. Cong, C. Tong, "Ultra-wideband robust RCS reduction with triangle-type AMC structure," *Radio engineering* 27, no. 2, vol, 403, 2018.

[19] Y. Saifullah, A. B. Waqas, G. M. Yang, F. H. Zhang, F. Xu, "4-bit optimized coding metasurface for wideband RCS reduction," *IEEE Access*, vol. 7, pp. 122378-122386, Aug. 2019.

[20] E. Ameri, S. H. Esmaeli, and S. H. Sedighy, "Low cost and thin metasurface for ultra-wide band and wide angle polarization insensitive radar cross section reduction," *Appl. Phys. Lett.* 112(20), 201601, 2018.

[21] F. Yang and Y. Rahmat-Samii, *Electromagnetic Band Gap Structures in Antenna Engineering* (UK Cambridge University Press, Cambridge, 2009.